\documentclass[12pt]{article}
\pdfoutput=1
\usepackage{overpic}
\usepackage{wrapfig}
\newcommand{\ie}{\emph{i.~e.},}
\newcommand{\eg}{\emph{e.~g.},}					

\newcommand{\phiphi}{\ensuremath{B_s^0 \to \phi \phi}}
\newcommand{\phikk}{\ensuremath{\phi \to K^+K^-}}
\newcommand{\jpsimm}{\ensuremath{ J\!/\!\psi \to \mu^+\mu^-}}
\newcommand{\jpsiphi}{\ensuremath{B_s^0 \to J\!/\!\psi \phi}}
\newcommand{\phikst}{\ensuremath{B^0 \to \phi K^{\ast 0}}}
\newcommand{\jpsikst}{\ensuremath{B^0 \to J/\psi K^{\ast 0}}}
\newcommand{\bhh}{\ensuremath{B_{(s)}^0 \to  h^+h'^-}}
\newcommand{\CP}{\ensuremath{\mathsf{CP}}}
\newcommand{\DGs}{\ensuremath{\Delta \Gamma_s}}
\newcommand{\Br}{\ensuremath{\mathcal{B}}}
\newcommand{\Or}{\ensuremath{\mathcal{O}}}
\newcommand{\bs}{\ensuremath{B_s^0}}
\newcommand{\pt}{\ensuremath{p_T}}
\newcommand{\lumifb}{fb$^{-1}$}

\def\trieste{Department of Physics, University of Trieste  \\
and INFN, Sezione di Trieste, ITALY}

\def\Title#1{\begin{center} {\Large #1 } \end{center}}
\def\Author#1{\begin{center}{ \sc #1} \end{center}}
\def\Address#1{\begin{center}{ \it #1} \end{center}}

\newcommand\pubblock{\rightline{\begin{tabular}{l} \\
         \end{tabular}}}
\newenvironment{Abstract}{\begin{quotation}  }{\end{quotation}}
\newenvironment{Presented}{\begin{quotation} \begin{center} 
             PRESENTED AT\end{center}\bigskip 
      \begin{center}\begin{large}}{\end{large}\end{center} \end{quotation}}

\begin{document}
\begin{titlepage}
\pubblock
\vfill
\Title{Charmless and Penguin Decays at CDF}
\vfill
\Author{ Mirco Dorigo (for the CDF Collaboration)}
\Address{\trieste}
\vfill
\begin{Abstract}
Penguin transitions play a key role in the search of New Physics
hints in the heavy flavor sector. During the last decade CDF has been
exploring this opportunity with a rich study of two--body charmless
decays of neutral $B$ mesons into charged final--state particles. 
After briefly introducing the aspects of this physics peculiar
to the hadron collision environment, I report on two interesting results: the
first polarization measurement of the \phiphi\ decay and the update of
the \bhh\ decays analysis.
\end{Abstract}
\vfill
\begin{Presented}
6$^{\textup{th}}$ International Workshop on the CKM Unitarity Triangle, University of Warwick, UK, 6--10 September 2010
\end{Presented}
\vfill
\end{titlepage}
\def\thefootnote{\fnsymbol{footnote}}
\setcounter{footnote}{0}

\section{Introduction: the penguin role}
Since the relevance of the penguin amplitudes was well established by the
CLEO experiment about ten years ago~\cite{cleo}, penguin--dominated processes
become more and more attractive despite their theoretical
complexity and experimental rareness.
The possibility to access to New Physics (NP)
through new virtual 
particles in the loops makes them an opportunity to be
exploited rather than a limitation.

Non--leptonic two--body charmless decays of neutral $B$ mesons,
with their sensitivity to penguin amplitudes,
play a major role. They can be used to determine the values of the CKM related
quantities, which may differ from the ones extracted from 
tree-level dominated processes, possibly indicating non-Standard Model
contributions.
Comparison between $\gamma$ determinations
from charmed tree--dominated decays and from charmless \bhh\ probes possible presence of 
NP in the penguin amplitude. Similarly, comparison of the mixing phase
between the tree--process \jpsiphi\   
and the penguin \phiphi\ charmless mode 
 could allow to
disentangle NP contribution in mixing and decay. 

Usually optimal sensitivity is obtained through 
a full tagged and time--dependent analysis 
(as the case of \jpsiphi\ above). 
However, this requires significant statistics of decays 
that are typically rare. 
Nevertheless, several strategies have been proposed that do not
require identification 
of production flavor and provide already some sensitivity
to NP contributions 
from measurements of averaged branching fractions, polarization
amplitudes and time--integrated 
\CP--asymmetries. 

The CDF experiment provides a joint access to large samples of
$B^0$ and $B^0_s$ mesons, which allows to study charmless decays into charged
final--state particles. This has been explored since the early 2000's with a pioneering and very rewarding
program (for instance~\cite{prog,BR1}); here, 
the latest measurements of the \phiphi\ decay and an overview of the
results of the \bhh\ analysis are reported, discussing finally some prospects. All conjugate modes are
implied and branching fractions are \CP--averages; for each result, the first uncertainty
is statistical and the second one is the systematic error.

\section{Charmless $B$ decays in hadron collisions}
CDF is a multipurpose solenoidal spectrometer with calorimeters and
muon detectors,
located at one of the two interaction points of the Tevatron
collider. The good performances of the Tevatron
(peak luminosity $\simeq 4 \times 10^{32}$ cm$^{-2}$ s$^{-1}$) and the
high data--taking efficiency allow to store on tape 
about 50 pb$^{-1}$ of data every week. 

Despite the higher production rate of $B$ mesons with respect to the
$B$--factories (\Or(10$^{3}$) more), the production cross section for a $b\bar{b}$ quarks pair at 1.96 TeV
(the Tevatron center of mass energy)
is a permille fraction of the total. Thus, a highly--discriminating 
signature is needed to distinguish relevant events of $b$--physics.
This is provided by the characteristic long lifetime of
$b$--hadrons:
they are typically produced with transverse momentum of a few GeV,
thus flying about 0.5 mm in the detector and resulting
in secondary vertices displaced from the $p\bar{p}$ collision point.
Triggering on those vertices is challenging. First,
it requires a high resolution tracking detector; this is given by 
double-sided silicon microstrips arranged in five cylindrical layers
and an open cell drift chamber with 
$96$ sense wires, all immersed in a 1.4 T solenoidal magnetic
field~\cite{CDF}.
Second, it needs to read out all the silicon detector (212~000 channels)
and do pattern recognition and track
fitting online. In CDF this is done by the Silicon Vertex Trigger in 25 $\mu$s
with a resolution of the track impact parameter (IP), 48 $\mu$m,
comparable with off-line measurements~\cite{Luciano}.
A sample enriched with $b$--flavor particles is then 
provided by a specific three--level trigger, based on the requirement
of two displaced tracks (\ie\ with IP larger than
\Or(120 $\mu$m)) with opposite charge.

\section{\phiphi\ Polarization}
The $\phiphi$  decay proceeds through a $b\rightarrow s\bar{s}s$ quark level
process, whose dominant SM diagram is the $b\rightarrow s$
penguin. The rich dynamics of decay of pseudoscalar meson 
in two vector particles involves 
three different amplitudes corresponding to the
polarization states. Hence, the \phiphi\ channel is
attractive to test the theoretical predictions for these polarization
amplitudes~\cite{BVV_th}, which have shown several discrepancies with measurements of
 similar penguin decays~\cite{BVV_exp}, raising considerable attention on the so--called
``polarization puzzle''.

The first evidence for the $\phiphi$ decay has been reported by CDF 
in 2005 with 8 events in a data sample corresponding to integrated
luminosity of 180 pb$^{-1}$~\cite{BR1}. 
An updated analysis recently improved the measurement of branching
fraction using 2.9 fb$^{-1}$ of data and allows the world's first
decay polarization
measurement~\cite{phiphi}. We reconstruct the \jpsiphi\ decay in the same dataset, and use this decay 
as a normalization for the branching ratio measurement and as a
control sample for the polarization analysis.

\begin{figure} 
\begin{center}
\begin{overpic}[width=0.45\columnwidth]{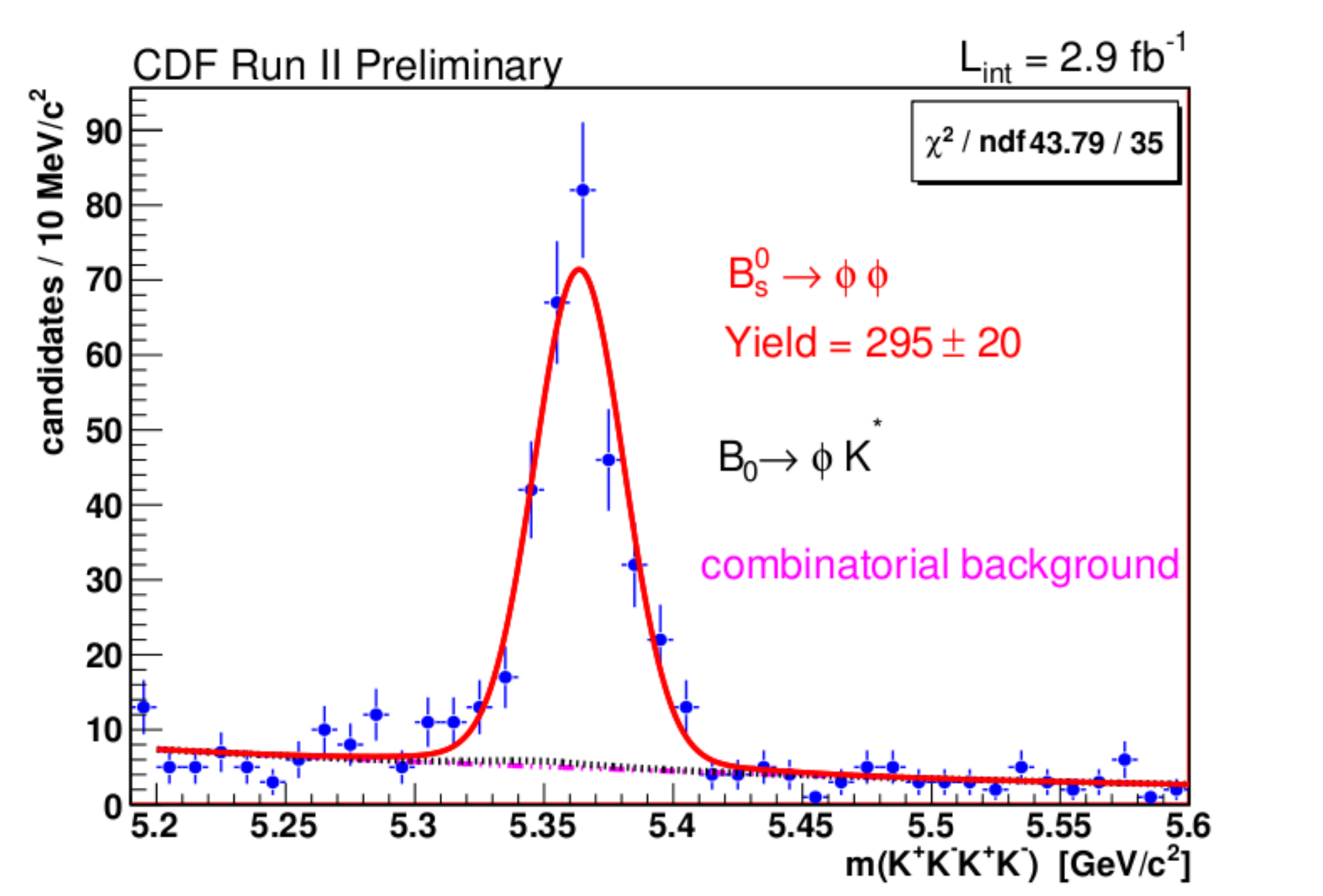}
\end{overpic}
\begin{overpic}[width=0.45\columnwidth]{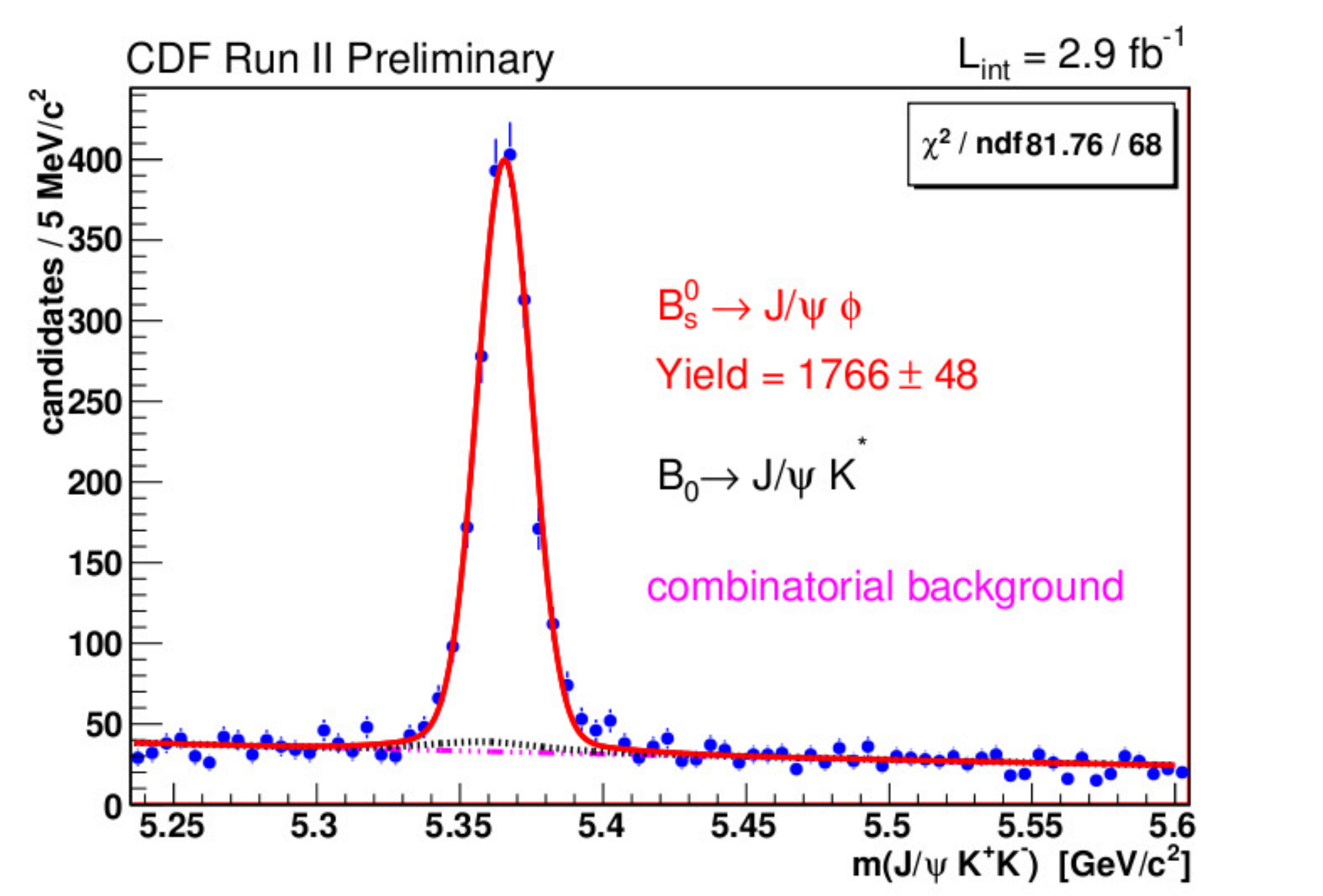}
\end{overpic}
\end{center}
\caption{\label{fig:peaks}
$m_{KKKK}$ (left) and $m_{J/\Psi KK}$ (right) distributions 
for \phiphi\ and \jpsiphi\ candidates with fit projections overlaid.}
\end{figure}

Signal candidates are reconstructed by detecting \phikk\ and \jpsimm\
decays and are formed by fitting four tracks to a common vertex. 
Combinatorial background is reduced by exploiting several
variables sensitive to the long lifetime and relatively
hard \pt\ spectrum of $B$ mesons.
The requirements on the discriminating variables are optimized by maximizing 
$\mathcal {S/\sqrt{S+B}}$, where
the signal ($\mathcal{S}$) is derived from a
Monte Carlo (MC) simulation and
the background ($\mathcal{B}$) is represented by appropriately normalized 
data sampled from the sideband mass regions.
Two sources of background are expected 
for both \phiphi\ and \jpsiphi:
a dominant and smooth combinatorial background
and a physics component, which 
is given by \phikst\ (\jpsikst) decays in the case of
\phiphi\ (\jpsiphi) and it is estimated by simulation not to exceed a
3\% fraction of the signal. Signals of $295\pm20$ events of \phiphi\ and $1766\pm 48$ events of \jpsiphi\ are obtained 
 by fitting the mass distributions (see
fig.~\ref{fig:peaks}).

The relative \phiphi\ decay rate is calculated using 
\begin{eqnarray}
\label{eq:BRphiphi}
\frac{\Br\left(\phiphi\right)}
     {\Br\left(\jpsiphi\right)} =
\frac{N_{\phi\phi}}{N_{\psi\phi}}
\frac{\Br\left(\jpsimm \right)}{\Br \left(\phikk \right)}
\frac{\epsilon_{\psi\phi}}{\epsilon_{\phi\phi}}\, \epsilon^{\mu}_{\psi\phi}\ ,
\end{eqnarray}
where $N_{\phi\phi}$ ($N_{\psi\phi}$) is the number of \phiphi\
(\jpsiphi) events, $\epsilon_{\phi\phi}/\epsilon_{\psi\phi}=0.939\pm 0.030 \pm 0.009$ is 
the relative trigger and selection efficiency extracted from
simulation; $\epsilon^{\mu}_{\psi\phi} = 0.8695 \pm 0.0044$, the
efficiency for identifying a muon, is obtained from data using inclusive
\jpsimm\ decays as a function of muon \pt. By using the known values for the branching
fractions \Br(\phikk), \Br(\jpsimm) and \Br(\jpsiphi) ~\cite{PDG} (updated to current values of 
$f_s/f_d$), we determine $\Br(\phiphi)=(2.40 \pm 0.21 \pm 0.27 \pm 0.82)
\times 10^{-5}$\footnote{The 
last uncertainty is given by the uncertainty in \Br(\jpsiphi).}
in agreement with the previous determination~\cite{BR1}.

We measure the polarization from the  angular distributions of decay products, expressed as a function of helicity angles,
$\vec{\omega}=(\cos\vartheta_1,\cos\vartheta_2,\Phi$). 
The total decay width is composed of three polarization amplitudes:
$A_0$, $A_\parallel$ and $A_\perp$. 
Taking the untagged decay rate integrated in time and
neglecting \CP-violating differences between $\bar{B}^0_s$ and $B^0_s$
both in mixing and in decay, the differential decay rate is a function of the
helicity angles and depends on the polarization amplitudes at $t=0$ and on
the light and heavy $B_s^0$ mass-eigenstate lifetimes, $\tau_{L}$ and $\tau_{H}$
respectively, as follows: 
\begin{equation}
\label{eq:decay_rate}
\frac{d^3\Gamma}{d\vec{\omega}} \propto \, \tau _{L}\big(
|A_0|^2f_1 + |A_\parallel|^2 f_2 
 + |A_0||A_\parallel|\cos\delta_\parallel f_5 \big) + \tau_{H}|A_\perp|^2f_3,
\end{equation}
where $\delta_\parallel=\arg (A_0^\star A_\parallel)$ and the
$f_i =f_i(\vec{\omega})$ are functions of the helicity angles only.

We perform an unbinned maximum likelihood fit to the reconstructed
mass $m$ of the $B_s^0$ candidates and the helicity
angles. The mass distribution provides discrimination between signal and background.
At this stage only the combinatorial component is considered as background,
accounting for the small contamination of other decays to the
systematic uncertainties.
Fixing $\tau_{L}$ and $\tau_{H}$ to the known
 values~\cite{PDG}, the polarization amplitudes are extracted
using directly Eq.~\ref{eq:decay_rate} as the angular
probability density function for the signal, corrected by
an acceptance factor given by a full MC simulation.
 The angular background is modeled
on sidebands data with polynomials and
fitted in the whole mass range.
The fitter is extensively tested and validated through
statistical trials; the robustness and reliability of the approach is
tested in data as well by measuring the polarization of the \jpsiphi\
in the sample used in the branching fraction update described above; the results, 
$|A_0|^2=0.534\pm0.019$ and
$|A_\parallel|^2=0.220\pm0.025$, are in very good agreement
with~\cite{jpsiphi}.
\begin{figure}[b]
\begin{overpic}[width=0.31\columnwidth]{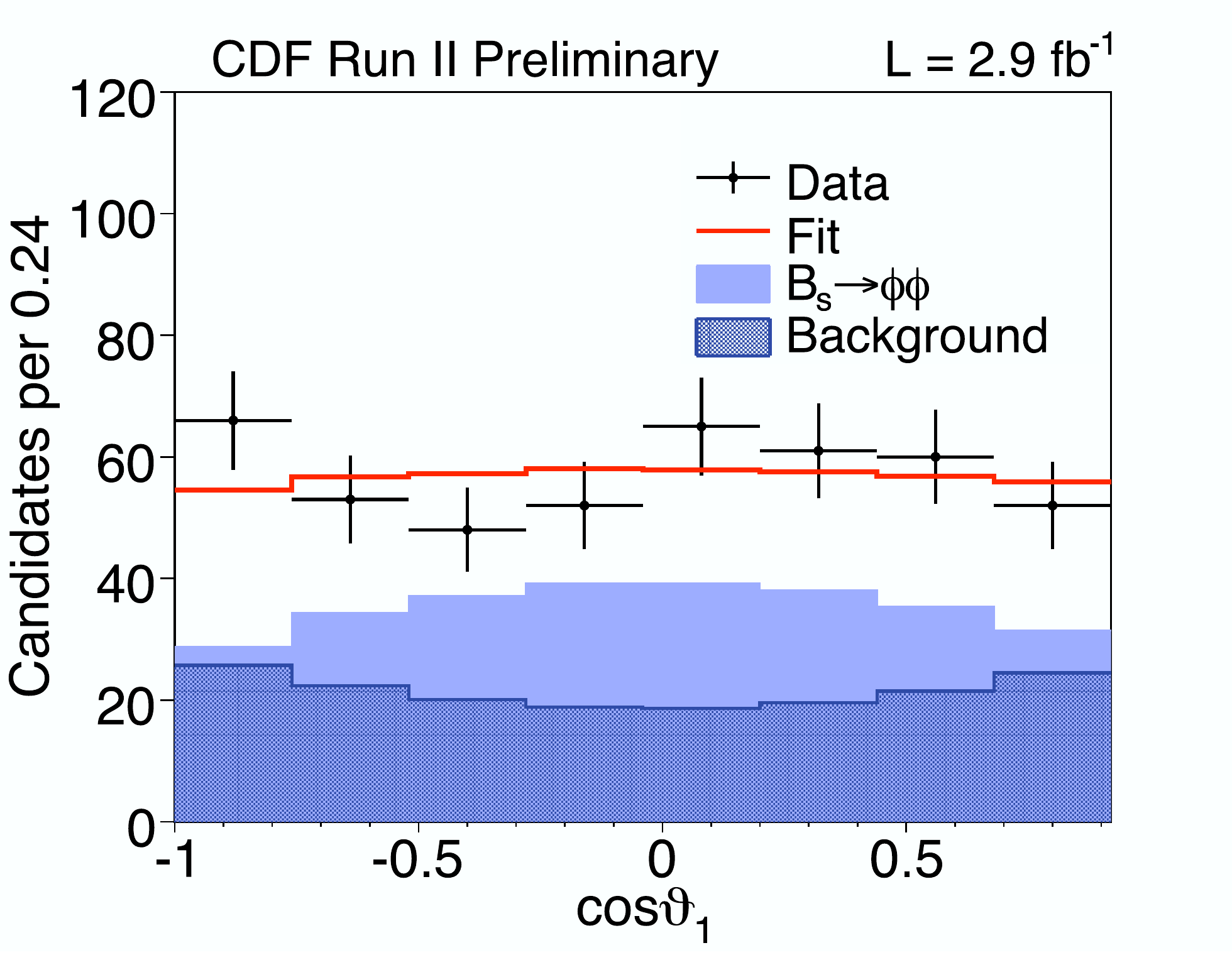}
\end{overpic}
\begin{overpic}[width=0.31\columnwidth]{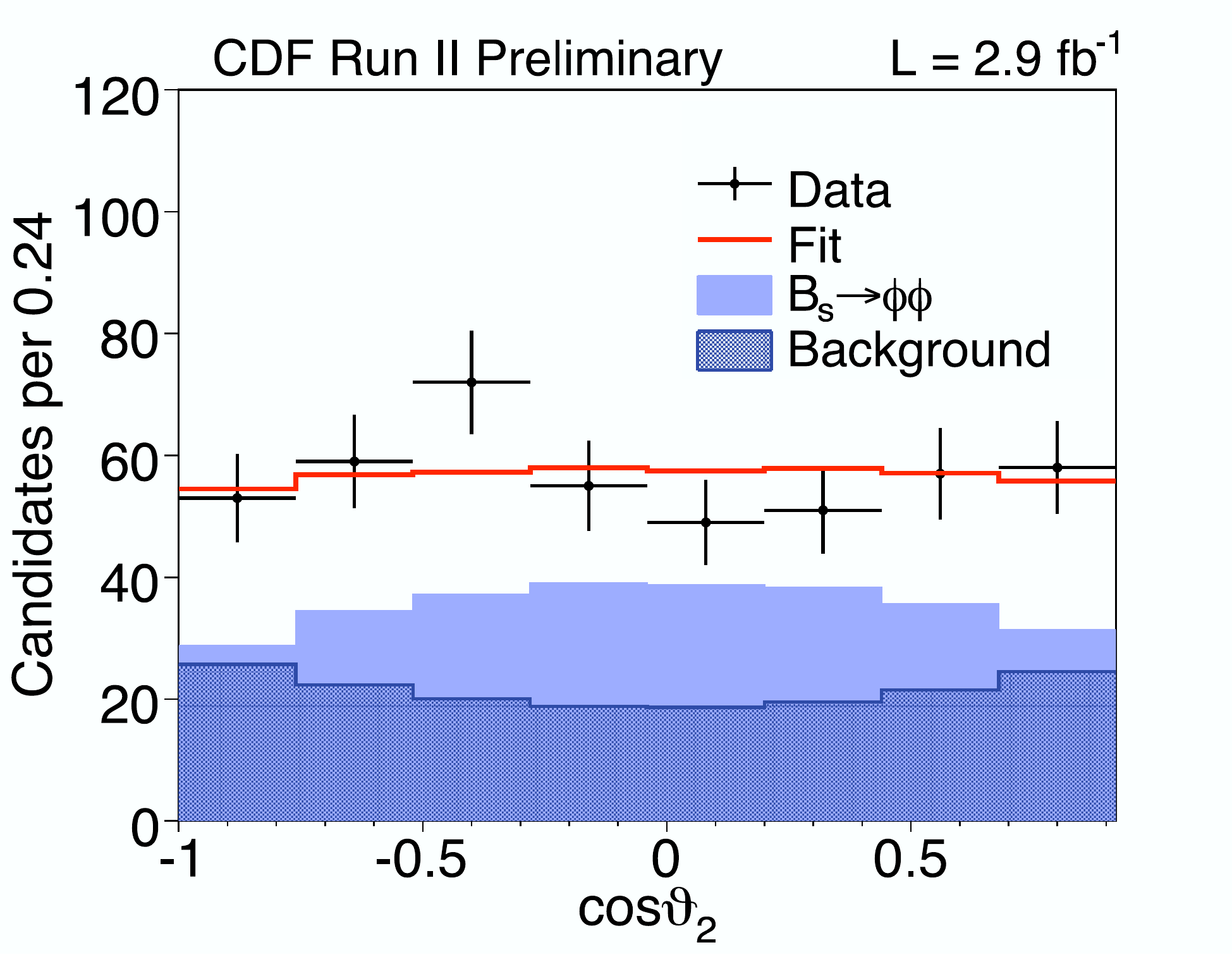}
\end{overpic}
\begin{overpic}[width=0.31\columnwidth]{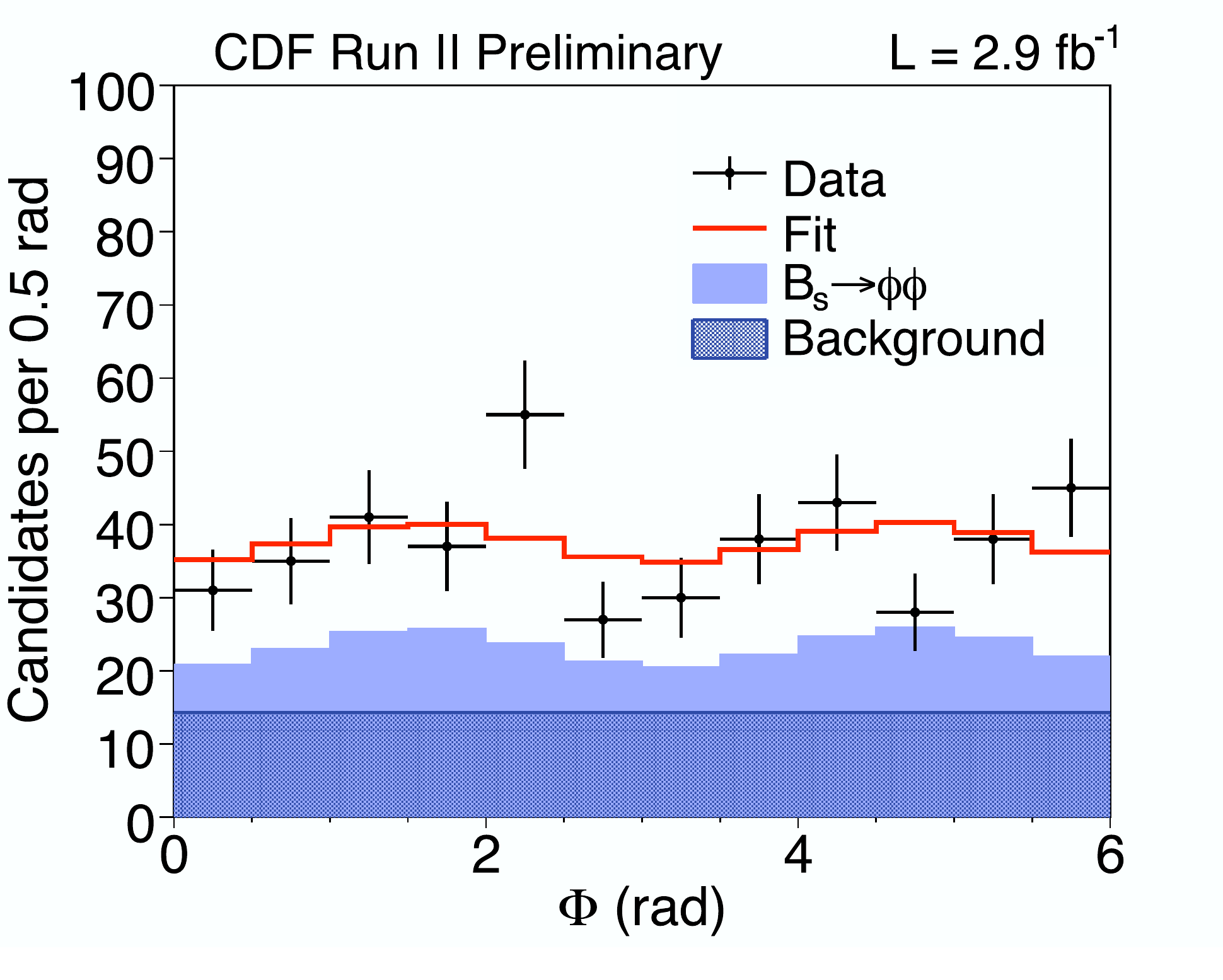}
\end{overpic}
\caption{\small Projections
 of the fit for \phiphi\ onto the helicity angles.}
\label{fig:ang}
\end{figure}

Several sources of systematic uncertainty are considered
 and only the main two are reported here.
We account for the physics background effects through simulated
samples. In addition to the \phikst\ decay, we consider
other two contributions, the $B^0_s \to \phi
f_0$ decay and the non--resonant $B^0_s \to \phi(K^+K^-)$:
all the background decays give an additional 1.5\% uncertainty on the polarization
estimates. Then, the biases (1\%) introduced by the time integration
of the decay rate are
examined with MC simulation. We check the tiny
impact (0.2\%) of our assumption of neglecting \CP-violating effects.

The angular
fit projections are shown in fig.~\ref{fig:ang}. 
The results of the polarization observables for the \phiphi\
candidates are $|A_0|^2=0.348 \pm 0.041 \pm 0.021$,
$|A_\parallel|^2=0.287 \pm 0.043 \pm 0.011$,
 $|A_\perp|^2=0.365 \pm 0.044 \pm  0.027$  and
$\cos\delta_\parallel = -0.91^{+0.15}_{-0.13} \pm 0.09$. 
The measured amplitudes result in a
smaller longitudinal fraction with respect to the naive expectation, 
$f_L=|A_0|^2 /(|A_0|^2+|A_\parallel|^2+|A_\perp|^2)=0.348\pm0.041\pm0.021$,
like previously found in
other similar $b \to s$ penguin decays~\cite{BVV_exp}.

\section{\bhh\ decays}
Two-body decays mediated by the $b \to u$ quark level transition have
amplitudes sensitive to $\gamma$ CKM angle and to NP too, being
affected by significant contributions from penguin transitions~\cite{bhh_th}. A variety of open
channels with similar final states characterize this transition: this
allow cancellation of many common systematic effects and provides
key information to improve effective models of low--energy QCD. 
In 1~fb$^{-1}$ of data, CDF observes four new modes of
these decays and has unique access to direct \CP\ violation in \bs\
and $\Lambda^0_b$ decays; in addition, our measurements of direct \CP\ violation
in the $B^0$ sector are competitive with the $B$--factories.
\begin{wrapfloat}{figure}{4}{0pt}
\includegraphics[width=0.38\columnwidth]{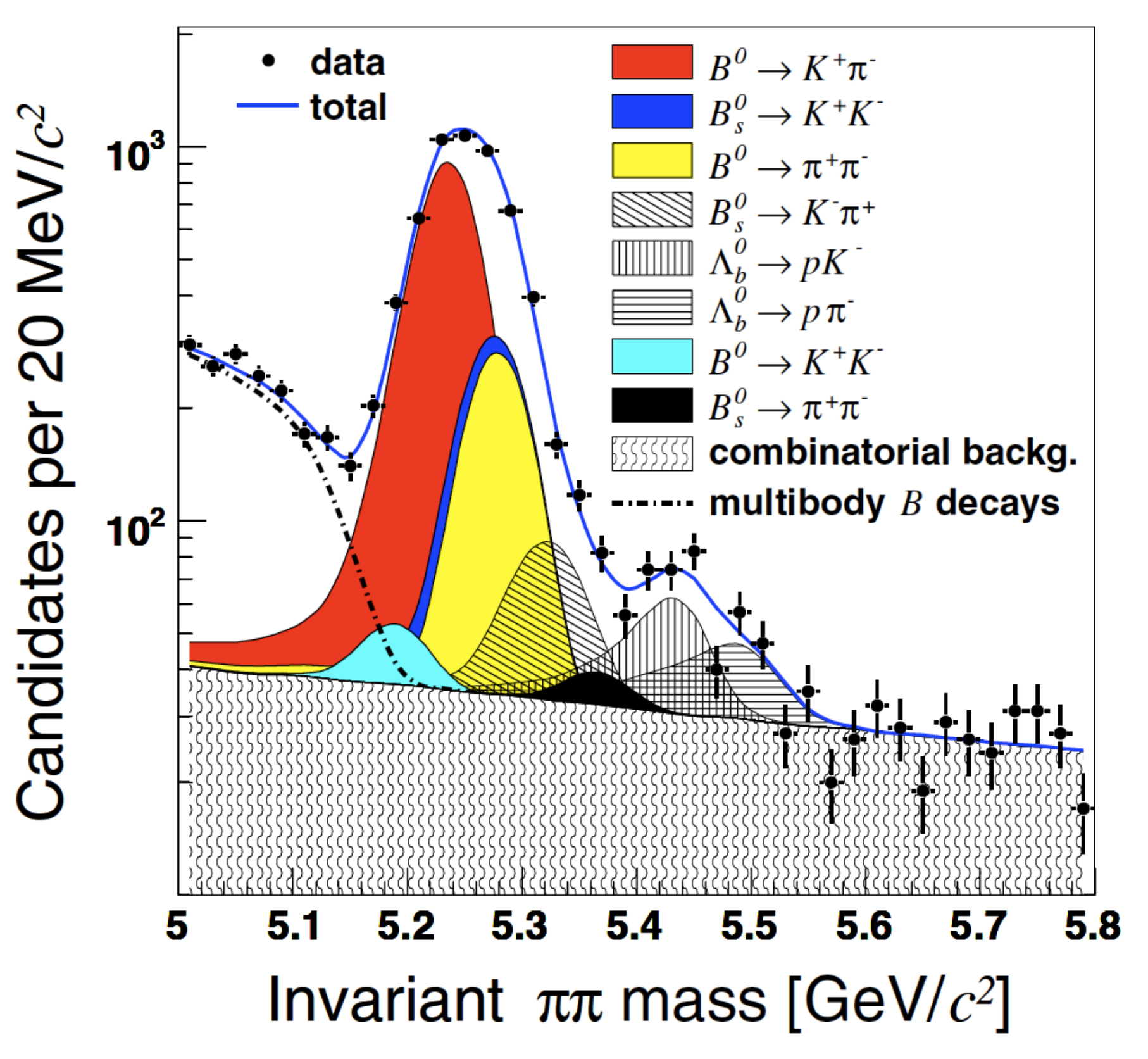}
\caption{\small Distribution of $\pi\pi$--mass for \bhh\ candidates
  with  fit projections overlaid.}
\label{fig:bhh}
\end{wrapfloat}

An optimized offline selection isolates a clear signal of roughly 7000 events over
a smooth background (fig.~\ref{fig:bhh}). 
A five-dimensional likelihood fit relying on kinematics differences between decays and particle
identification from the measurement of specific ionization in the
drift chamber allow statistical determination of the individual
contributions.
Event fractions for each channel are corrected for trigger and selection efficiencies
(from simulation and data) to extract the decay--rates. We report the first observation
of the decays $\bs\to K^-\pi^+$, $\Lambda^0_b \to p \pi^-$, $\Lambda^0_b \to p K^-$, 
 and world-leading measurements of (upper
limits on) the $\bs\to K^+K^-$ ($\bs\to \pi^+\pi^-$) branching
fractions: in unit of $10^{-6}$, $\Br(\bs\to K^-\pi^+)=5.0\pm 0.7 \pm 0.8$,
 $\Br(\Lambda^0_b \to p \pi^-)=3.5\pm 0.6\pm 0.9$,
 $\Br(\Lambda^0_b \to p K^-)=5.6\pm 0.8\pm 1.5$,
  $\Br(B^0 \to \pi^+\pi^-)=5.02 \pm 0.33 \pm 0.35$,
  $\Br(\bs \to K^+K^-)=24.4 \pm 1.4 \pm 3.5$,
 and $\Br(\bs\to \pi^+\pi^-)<1.2$,
 $\Br(B^0\to K^+K^-)<0.7$, both at 90\% C.L. The measured ratio 
$\Br(\Lambda^0_b \to p \pi^-)/\Br(\Lambda^0_b \to p K^-)=0.66\pm
0.14\pm 0.08$ shows that significant penguin contributions compensate
the Cabibbo (and kinematic) suppression expected at tree level. 
We also report first measurements
of the following direct \CP--violating asymmetries: $A_\CP(\bs\to K^-\pi^+) = (39\pm15\pm8)\%$,
$A_\CP(\Lambda^0_b \to p \pi^-) = (3\pm 17\pm 5)\%$, and
$A_\CP(\Lambda^0_b \to p K^-) = (37 \pm 17 \pm 3)\%$. Direct \CP\ violation
in the $B^0\to K^+\pi^-$ mode is measured as $A_\CP(B^0\to K^+\pi^-) =
(-8.6\pm 2.3 \pm 0.9)\%$, consistent with the final $B$--factories
results. 
The analysis of the 6 fb$^{-1}$
sample is now in progress. We expect to have more than 16 000 \bhh\ decays where new
modes could be observed and \CP-violating asymmetries measured with doubled precision.

\section{Conclusions and Outlook}
Since the beginning of Run II, CDF is leading a rich program on
\bs\ meson physics, whose two--body
charmless decays are a key part, given their sensitivity to NP through rare penguin transitions.
We report the update of the branching fraction and the first
polarization measurement for the \phiphi\ decay, which
provides useful information on
the puzzling scenario of $B$ to vector--vector decays. 
Having a self-conjugate final state, the \phiphi\ can be used to measure the \bs\ decay width
difference (\DGs), and it is sensitive to the \CP\ violation in the decay 
and/or mixing, 
supplementing the
analogous measurements in tree-dominated \jpsiphi\ decay.
In the \bhh\ analysis, $B^0_s \to K^-\pi^+$, $\Lambda^0_b \to p \pi^-$, 
and $\Lambda^0_b \to p K^-$ decays were newly observed and their 
branching fractions and \CP-asymmetries measured.
Updates beyond 1~fb$^{-1}$ of this analysis will provide stringent
model--independent constraints on non--SM contributions and exploration of
time--dependent \CP--asymmetries, increasing our knowledge of
the $\gamma$ angle. 
CDF  collected to date 8~\lumifb\ of physics-quality data, which will
reach 10~\lumifb\ by October 2011. 
Additional 6~\lumifb\ will be collected if the proposed three-year extension will be funded.

\small{

} 
\end{document}